# COST EFFECTIVENESS STATISTIC: A PROPOSAL TO TAKE INTO ACCOUNT THE PATIENT STRATIFICATION FACTORS

*C. D'Urso, IEEE senior member*

**ABSTRACT**

The solution here proposed can be used to conduct economic analysis in randomized clinical trials. It is based on a statistical approach and aims at calculating a revised version of the incremental cost-effective ratio (ICER) in order to take into account the key factors that can influence the choice of therapy causing confounding by indication. Let us take as an example a new therapy to treat cancer being compared to an existing therapy with effectiveness taken as time to death. A challenging problem is that the ICER is defined in terms of means over the entire treatment groups. It makes no provision for stratification by groups of patients with differing risk of death. For example, for a fair and unbiased analysis, one would desire to compare time to death in groups with similar life expectancy which would be impacted by factors such as age, gender, disease severity, etc. The method we decided to apply is borrowed by cluster analysis and aims at *(i)* discard any outliers in the set under analysis that may arise, *(ii)* identify groups (i.e. clusters) of patients with "similar" key factors.

**INTRODUCTION**

The value of a new drug or therapy is often evaluated in terms of Incremental Cost Effectiveness Ratio (ICER) [1]. It is an equation used commonly in health economics to provide a practical approach to decision making regarding health interventions, in fact within a trial of two interventions it is the measure primarily used to compare the cost-effectiveness of the experimental treatment relative to the control treatment. The equation for ICER is:

$$ICER = \frac{\overline{C_e} - \overline{C_c}}{\overline{E_e} - \overline{E_c}},$$

where $\overline{C_e}$ and $\overline{C_c}$ are the mean costs, and $\overline{E_e}$ and $\overline{E_c}$ are the mean effects for the experimental and control treatments, respectively.

In order to take into account characteristic key factors of the therapy (e.g. age, LVEF and NYHA in a ICD therapy [2]), we consider the generic set P of n patients each described by m variables (key factors) as a data table $P(n, m)$ with n rows and m columns. Our goal can be stated as follows:

**Goal**. Identify k subsets (called $SP_j$) of the set P containing $n_j$ elements with $\sum_1^k n_j \leq n$, such that each of them contains patients with similar key factors, and such that: $\sum_1^k n_j + n_{out} = n$, with $n_{out}$ the numbers of outliers discarded by the method. The calculation of the index $ICER_j$ on the elements of each group of patients must lead us to construct the overall incremental cost effectiveness ratio (ICER).

Although we could assume some statistical properties of the distributions involved, in the following we will work under the following hypothesis:

**Hypotesis**. No assumption about prior distribution or other parametric choices will be considered.

**RELATED WORK**

Methods for describing the distributional properties of ICER statistics have been reported in the past, however, they are challenging even before the notion of stratification is added. In general, the distribution of the ratio of two statistics (in this case, the ratio in the difference of two means) is complicated. Interesting approaches such as Fieller's theorem have been suggested. More successful have been approaches base on the bootstrap [3]. However the work of Abadie and Imbens [4] suggests more caution should be exercised before approaches based on patient matching and stratification are embarked upon.

**RATIONALE OF THE METHOD**

An outlier is an observation that, as atypical or erroneous, deviates significantly from the behavior of other data, with reference to the type of analysis considered [5, 6, 7]. In our problem there could be among the patients a subset with some characteristics (i.e. key factors) very dissimilar to the others that could negatively affect the economic analysis in randomized clinical trials.
The method proposed exploits the theories of cluster analysis, and as a first step it identifies those points of the set under analysis that have "distance" from the centroid (in terms of standard deviations) more than a specific threshold away. As an example to make clear the idea, in figure 1 is depicted the Mahalanobis distance from the centroid of each point in a certain set of data. The potential outliers are those more distant from the centroid. The Mahalanobis distance is one of the possible distance that can be considered, it has the important characteristic that it takes into account the context of the data that is the relation between data. The rational is the same for the more common euclidean distance as well.

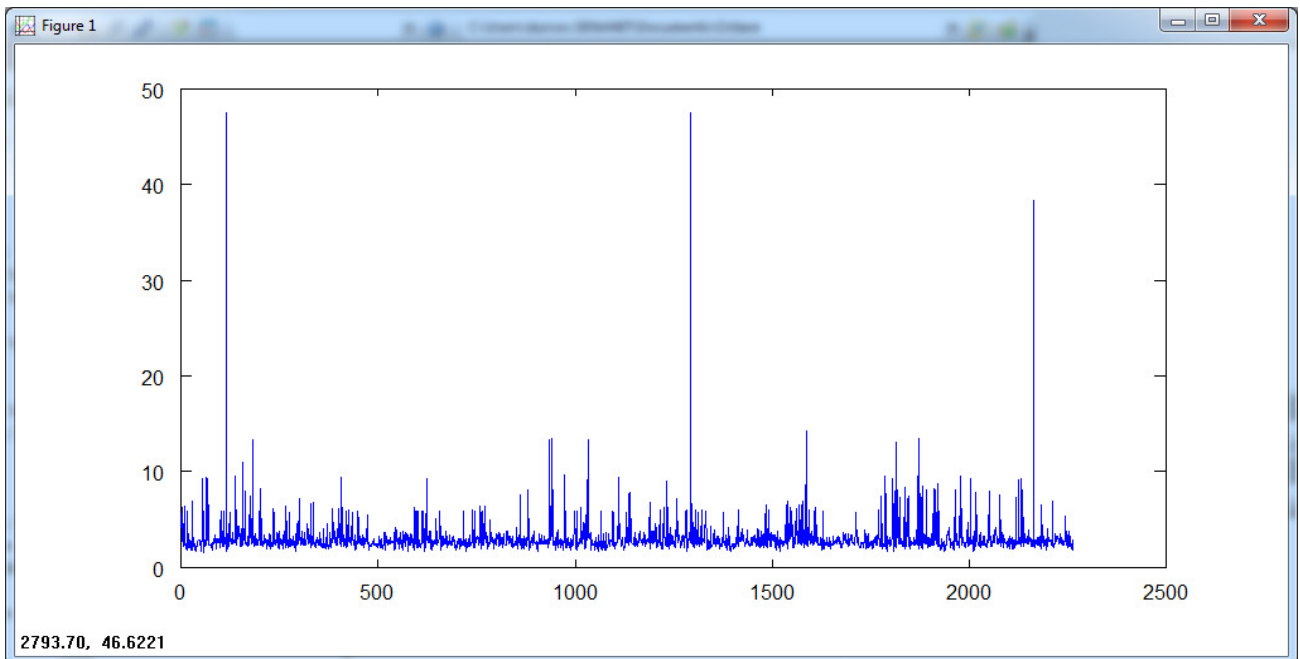

Figure 1 - Distance from each point in a set to the centroid (the spikes are potential errors in the set)

The cluster analysis applied to the set of patients data identifies a certain number (say *k*) of groups. In each of these group it will be calculated the incremental cost effective ratio. At this point the analysis applied may provide a situation similar to that depicted in the Figure 2.

The clusters obtained in this way are composed intrinsically of a homogeneous set of individuals and contain both patients belonging to the experimental treatment group and patients belonging to the control treatment group. Thus for each of these clusters it is possible to calculate the index $ICER_j$, and then calculate the overall ICER in accordance with the formula provided in the following (see Step 3).

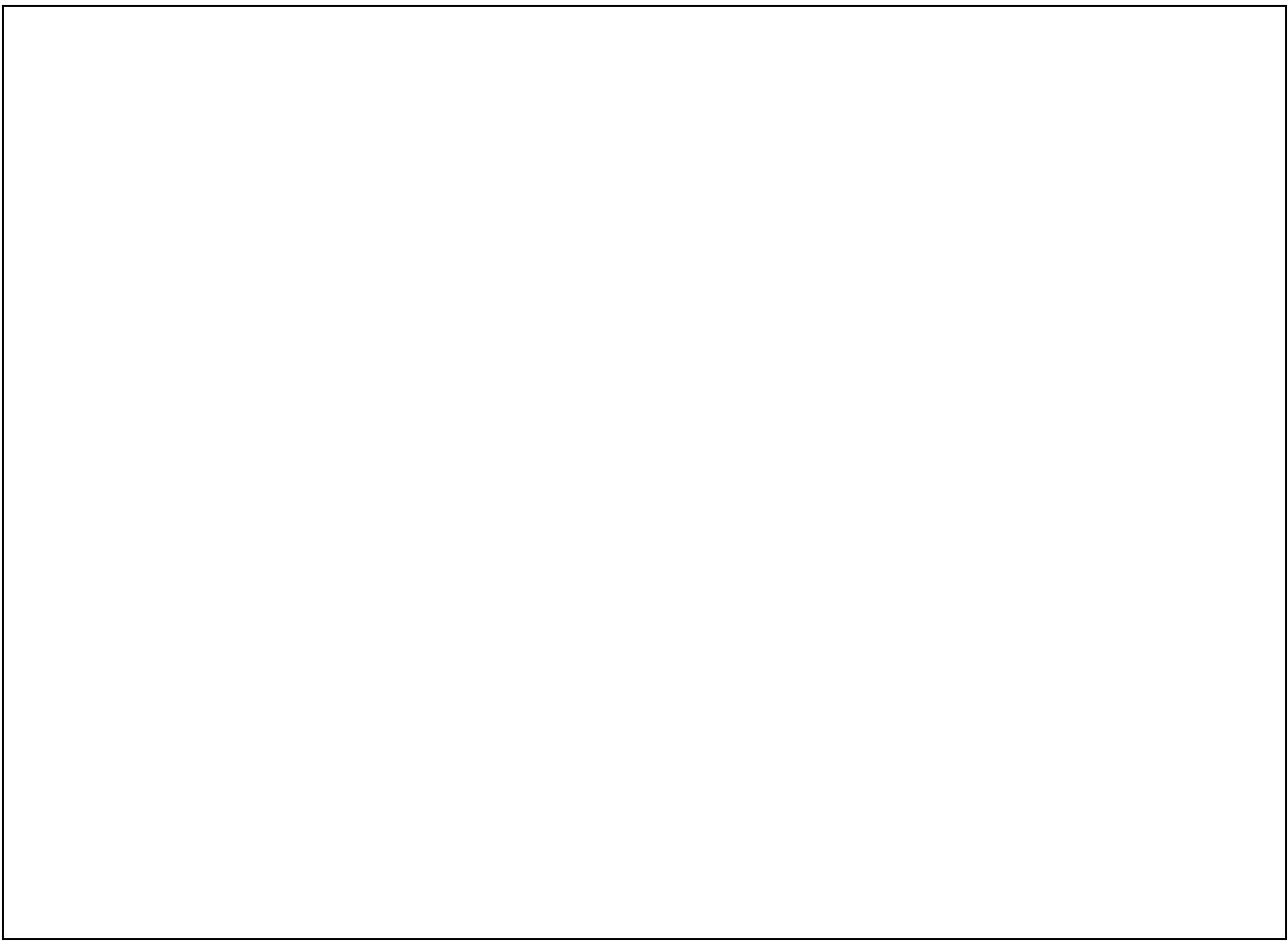

Figure 2 - Representation of a typical scenario after the application of cluster analysis to the data of the patients involving in clinical trials.

## SOLUTION

Here it is proposed a method based on cluster analysis. In particular we try to automate the calculation by means of classification of patients in a set of clusters in order to achieve the goal.

The method is composed of three sequential steps described in the following (the diagram is provided in Figure 3):

STEP ONE (cluster analysis)

First of all the data related to the patients are stored in an array P of n rows (total number of the patients involved in the clinical trial) and m columns (number of key factors). We assume that from a particular attribute of each patient (e.g. the identification code) we can distinguish if he belongs to the experimental treatment group or to the control treatment group.

The algorithm called DBSCAN is applied to the objects in P. It is a data clustering algorithm proposed by Martin Ester, Hans-Peter Kriegel, Jörg Sander and Xiaowei Xu in 1996 [8]. It is a density-based clustering algorithm because it finds a number of clusters starting from the estimated density distribution of corresponding nodes. DBSCAN is one of the most common clustering algorithms and also most cited in scientific literature (see appendix for the code provided in the Octave software environment [9]).

After the application of the algorithm, the objects (i.e. the rows of the array) are classified in k clusters plus a set of outlier data (cluster labeled '-1') that represents the patients with characteristics too different from the others as to suggest that they have been wrongly chosen for the trial (see Figure 2).

STEP TWO (calculation of each ICER$_j$)

Once we got the k groups of patients we can calculate k indexes ICERj distinguishing between the individuals belonging to the experimental treatment group and those belonging to the control treatment group, as follows:

$$ICER_j = \frac{C_j^e - C_j^c}{E_j^e - E_j^c}, j = 1..k$$

where:

$k$ are the clusters identified in step 1

$C_j^e$ and $C_j^c$ are the mean costs, and $E_j^e$ and $E_j^c$ are the mean effects for the experimental and control treatments, respectively.

STEP THREE (calculation of the overall ICER)

The final step consists in the calculation of the overall incremental cost effectiveness ratio based on a weighted average of contributions from the k ratios obtained in step two.

$$ICER = \sum_1^k \frac{n_j}{n} ICER_j$$

where:

$n = |\{Patients\ in\ the\ control\ goup\}| + |\{Patients\ in\ the\ experimental\ group\}| = n^c + n^e$

$n_j = n_j^e + n_j^c$, number of patients in the cluster j

$n_{out}$ number of patients discarded as potential outliers

$n = \sum_1^k n_j + n_{out}$ , number of patients involved in the clinical trial

In the following diagram is represented the logical flow of the method.

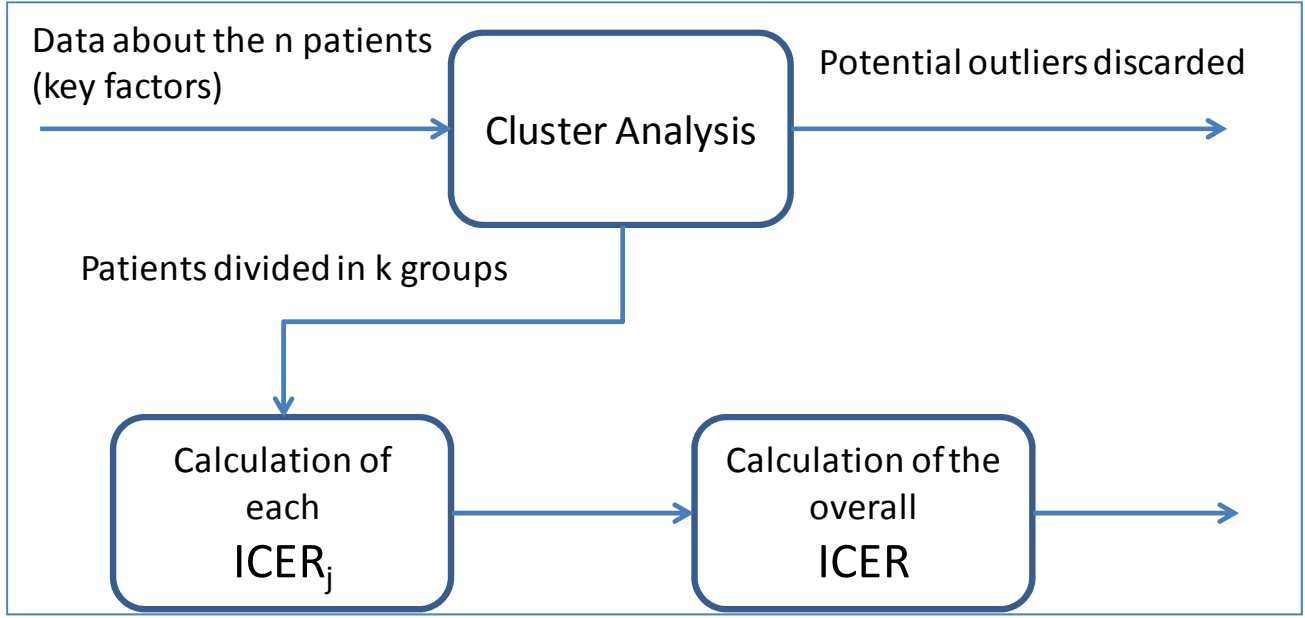

Figure 3 - Logical flow of the proposed solution.

**CONCLUSION**

The method proposed takes into account the effect of strata in a population of patients. It uses the key factors associated to each patients as inner properties in order to group them and apply the computation separately for each homogenous group. No assumption about prior distribution or other parametric choices has been considered. The descriptive parameters of the new overall ICER can be written as a function of the corresponding descriptive measures of each index ICER$_j$:

$$mean(ICER) = \sum_1^k \frac{n_j}{n} mean(ICER_j)$$

$$var(ICER) = \sum_1^k (\frac{n_j}{n})^2 var(ICER_j)$$

Therefore the variability of overall ratio can be expressed in terms of variability of the k indexes ICER$_j$ (see reference [10] for a comparison of four methods of confidence intervals computation for the not stratified version of ICER).

The action of disregarding anomalous data (see step 1 of the process and the notion of outliers) introduces a particular robustness in the proposed method. First of all we observe that clustering the data as the first step of the method allows analysis of homogeneous group of patients minimizing the overall effect of the observational nature of the data (i.e. non-random). We can state that such an algorithm is able to create subsets within a patient population that can provide more detailed information about how the patient will respond to a given drug. Furthermore missing data or anomalous data can be distinguished and labeled as outliers in the first step of the process, and besides if there are some patients with censored data it is very likely that they are grouped in the same cluster.